\documentclass{ws-mpla}
\usepackage[super]{cite}
\usepackage{graphicx}
\usepackage{subfigure}
\begin{document}

\markboth{N. Q. Lan, et al.} {Self-Interacting Dark Matter}


\title{3-3-1 Self Interacting Dark Matter and the Galaxy Core-Cusp problem}

\author{\footnotesize Nguyen Quynh Lan, Grant J. Mathews, Lara Arielle Phillips, Miguel A. Correa}

\address{Center for Astrophysics, Department of Physics, University of Notre Dame, Notre Dame, IN 46556, U.S.A.\\
lnguyen3@nd.edu, gmathews@nd.edu, Lara.A.Phillips.127@nd.edu,mcorrea2@nd.edu  }

\author{In-Saeng Suh}

\address{Center for Astrophysics, Department of Physics, and Center for Research Computing, University of Notre Dame, Notre Dame, IN 46556, U.S.A \\
isuh@nd.edu}

\author{Jared. W. Coughlin}

\address{NCSA, University of Illinois, Champaign, IL 61801, USA\\
 jcoughl3@illinois.edu}
\maketitle

\pub{Received (Day Month Year)}{Revised (Day Month Year)}

\begin{abstract}
The core-cusp problem remains as a challenging discrepancy between observations and simulations in the standard $\Lambda$CDM model for the formation of galaxies. The problem is that $\Lambda$CDM simulations predict a steep power-law mass density profile at  the center of galactic dark matter halos. However, observations of dwarf galaxies in the Local Group reveal a density profile consistent with a nearly flat distribution of dark matter near the center. A number of solutions to this dilemma have been proposed. Here, we summarize investigations into the possibility that the dark matter particles themselves self interact and scatter. Such self-interacting dark matter (SIDM) particles can smooth out the dark-matter profile in high-density regions. We also review the theoretical proposal that  self-interacting dark matter may arise as an additional Higgs scalar in the 3-3-1 extension of the standard model. We present new simulations of galaxy formation and evolution for this formulation of self-interacting dark matter. Current constraints on this  self-interacting dark matter are then summarized.

\end{abstract}


\section{Introduction}

\noindent Understanding the nature of most of the  matter in the
universe remains a challenge  in modern cosmology.\cite{mt00}   $\Lambda$CDM models with a mixture of
roughly 25\% collisionless cold dark matter, such as WIMPs, axions, massive neutrinos,
etc., that interact through the weak and
gravitational forces, plus about 70\% cosmological constant (or vacuum dark energy density)
match current observations of the cosmic microwave background and large
scale structure with remarkable
accuracy.\cite{bah99,wan00}  Indeed, it is now believed that only a fraction of
the present total matter can be made of ordinary baryons, while most of the mass-energy content of the universe has an unknown, nonbaryonic origin.\cite{pos01}  

Evidence for the existence of cold dark matter is derived from observations of the cosmic microwave background (CMB), galaxy clusters, weak gravitational  lensing, and the Lyman-$\alpha$ forest.\cite{Bertone18}  For the most part these observation  agree with the predictions of the $\Lambda$CDM model.
 However, it is now widely appreciated\cite{Tulin17}   
that conventional models of collisionless cold dark matter can
lead to problems with regard to galactic structure. 

Indeed, the $\Lambda$CDM model is
only able to fit the observations on large scales ($^>_\sim 1 $kpc).
In particular, high-resolution $N$-body $\Lambda$CDM  simulations invariably  result in a central
singularity within the dwarf galaxies of the galactic halo.\cite{ghi} 
This is know as the {\it Core-Cusp Problem}. This problem and a  number of other inconsistencies in the $\Lambda$CDM model have been  reviewed in a number of publications 
~\cite{Tulin17,dav01,bul00,Mathews14} . 
In addition to the core-cusp problem there are several other difficulties in $\Lambda$CDM simulations.\cite{Mathews14}. 

Numerical $\Lambda$CDM simulations generically  predict a hierarchical clustering  scenario with an ever increasing number count for smaller  sized proto-galactic halos.  Although the distribution of normal-sized galaxies in most simulations seems reasonable, the predicted number of dwarf galaxies\cite{Mateo98} is orders of magnitude lower than that expected from simulations \cite{Moore99}.  For example there are  currently less than  200   dwarf galaxies possibly associated with  the Local Group, and only  a subset  of those are associated with the Milky Way \cite{Mateo98}.  On the other hand,   dark matter simulations predict \cite{Moore99, Mathews14} $^>_\sim  500$ dwarf satellites orbiting the Milky Way alone. It seems unavoidable that  $\Lambda$CDM simulations predict the existence of a much larger number of  satellite dwarf galaxies  than is observed.  This is the missing satellites problem.

Another problem with $\Lambda$CDM simulations is 
the {\it  too-big-to-fail problem}~\cite{Boylan-Kolchin11, Boylan-Kolchin12a,Mathews14}
that the dark matter masses of the dSphs are significantly
lower than that expected from  $\Lambda$CDM simulations. The predicted  dense bright satellites are not observed.  

In the present work we restrict our discussion to the implications of the core-cusp problem, keeping in mind that the solution to this problem will also impact and possibly solve the other discrepancies as well.\cite{Mathews14}
We summarize constraints on the properties of  SIDM based upon a solution to the core-cusp problem.  As a particular example, we constrain the 3-3-1  model with a Higgs triplet as candidates for  self-interacting dark matter.\cite{ll}  We also constrain this model with simulations of galactic structure and the cosmic microwave background.

\begin{figure}[htp]
\begin{center}
\includegraphics[height=6cm]{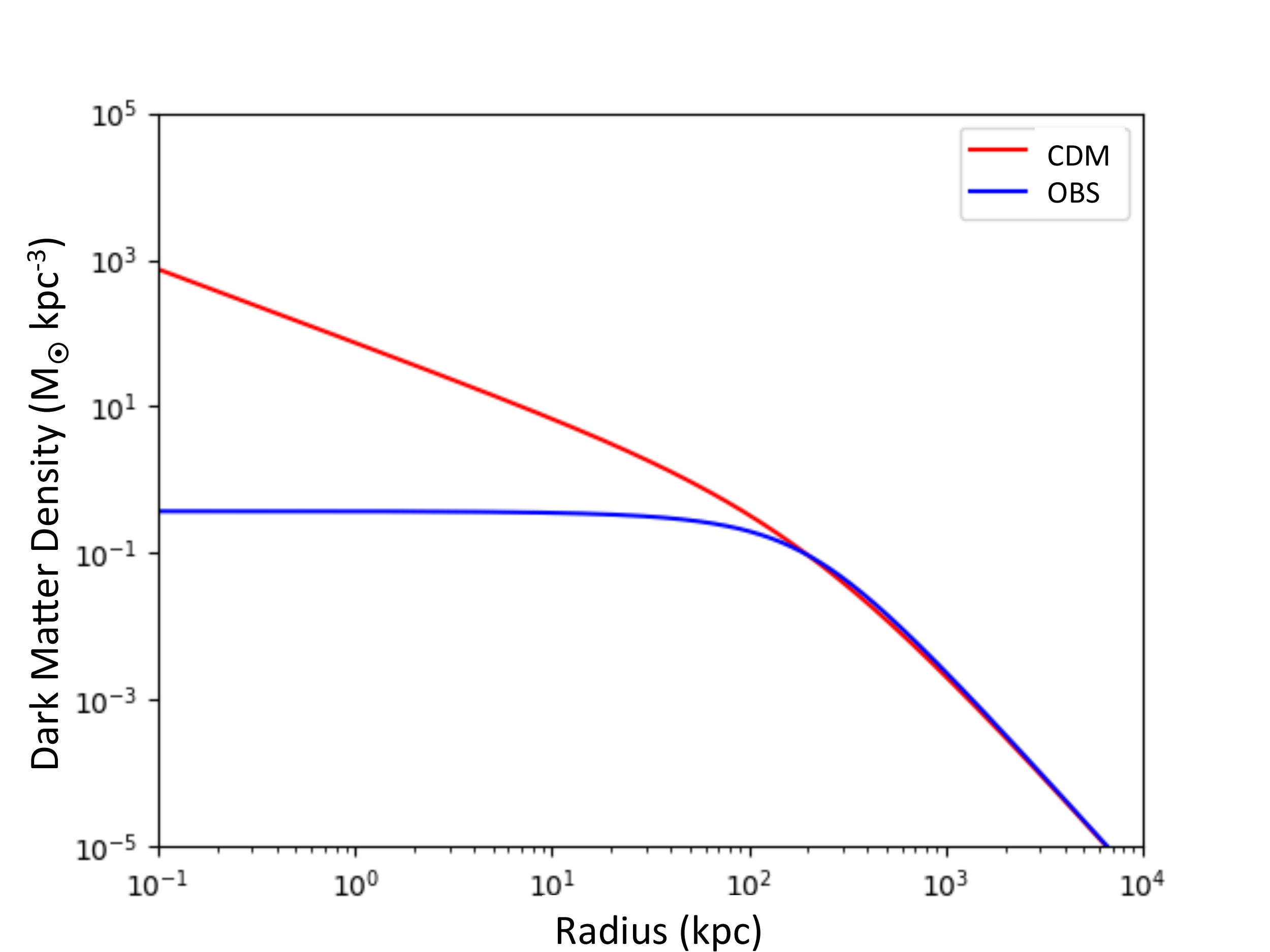}
\caption{Illustration of the core-cusp problem.  Blue line indicates the  observed dwarf-galaxy DM profiles while the red line shows the results of collisionless cold dark matter simulations.}
\label{fig:1}
\end{center}
\end{figure}

The core-cusp problem arises because the predicted mass density profile\cite{Dubinski91, Navarro96, Navarro97}  for the central region of CDM halos scales approximately as $\rho_{\rm dm} \sim r^{-1}$.  However, observed rotation curves of disk galaxies, dwarf galaxies, and low surface-brightness galaxies  exhibit a nearly constant inferred dark matter  density profile.\cite{Flores94,Moore94,Moore99}  This is most apparent\cite{Burkert95,McGaugh98}  in dwarf and low surface bright-ness galaxies because they can be  dominated by dark matter. 
In particular, CDM  haloes have spherical average density profiles that can be fit, for example,  with the NFW profile.\cite{Navarro96}
\begin{equation}
\rho_{CDM}(r) = \rho_0 \frac{r_s^3}{r(r+r_s)^2}~~,
\label{eq:1}\end{equation}
 where $r_s$ is a gravitational scattering length.  On the other hand,  the observed density profiles are better fit with a profile of the form:\cite{Burkert95}
\begin{equation}
\rho_{OBS}(r) = \rho_0 \frac{r_s^3}{(r+r_s)(r^2+r_s^2)}~~.
\label{eq:2}
\end{equation}
 A schematic illustration of the difference between these distributions is shown in figure  \ref{fig:1}.

Of interest to this review is that a possible way to
avoid these problems is to hypothesize 
self-interacting dark matter 
models as a well motivated alternative to the standard $\Lambda$CDM model \cite{ss}.

 It is a well-accepted fact that plausible
candidates for dark matter are elementary particles. The key
property of these particles is that  they  must  have a weak
scattering  cross-section and presently be non-relativistic. In response to the original work of Spergel-Steinhard, \cite{ss} many
follow-up studies have been made~\cite{ost,mcd} 

However, a number of authors\cite{ft,ppf,l97} have
proposed models in which a specific scalar singlet is introduced into the
standard model that satisfies
the constraints on self-interacting dark matter properties.

\section{3-3-1 Models for Self-Interacting Dark Matter }

 The standard model of particle physics (SM) offers no options for either dark matter or SIDM. The first gauge
model for  SIDM was proposed by Fregolente and Tonasse\cite{ft} in
the  3-3-1 model. In that model, one has to propose the existence of  {\it exotic leptons}
 to keep three triplets within the Higgs sector. 
 
 The
3-3-1 models were originally proposed based upon an independent
motivation\cite{ppf}. That is, these models have  intriguing
features such as
 they are anomaly free only if the number
of families $N$ is a multiple of three. If one also adds the
condition of QCD asymptotic freedom is valid only if the
number of families of quarks is less than five. It follows
that $N$ is equal to 3.

 A subject that has not been deeply explored, however, is the constraint imposed by galactic structure on this form of  dark matter.
 In Ref.~[\refcite{ll}] it was noted  that
 the   3-3-1 model with right-handed (RH)
neutrinos\cite{flt, lanlong} furnishes a good candidate for self-interacting dark
matter. 

The main properties that a good dark matter candidate must
satisfy are stability and neutrality. Therefore, one can consider the
scalar sector of the model, more specifically to the neutral
scalars. One must then establish that they are  stable and  can satisfy the constraints\cite{ss} on self-interacting dark matter. 
In addition, one requires that such dark matter particles have the correct contribution to the closure density of the universe ($\Omega_{DM} \sim 0.25$). On the
other hand, since this dark matter particle is not imposed
arbitrarily to solve this specific problem, one must check that the
required  cosmological parameters do not violate  other bounds
of the model.

 Under assumption of the discrete parity symmetry  $\chi \leftrightarrow -
\chi$, the most general potential for the Higgs fields $\eta, \rho, \chi,$ in the 3-3-1 scenario can be written in the
following form\cite{l97} 

\begin{eqnarray} 
V(\eta,\rho,\chi)&=&\mu^2_1 \eta^+
\eta +
 \mu^2_2 \rho^+ \rho +  \mu^2_3 \chi^+ \chi +
\lambda_1 (\eta^+ \eta)^2 + \lambda_2 (\rho^+ \rho)^2  \nonumber \\
& +&
\lambda_3 (\chi^+ \chi)^2 \nonumber 
+ (\eta^+ \eta) [ \lambda_4 (\rho^+ \rho) + \lambda_5 (\chi^+
\chi)] + \lambda_6 (\rho^+ \rho)(\chi^+ \chi) \nonumber \\
&+&
\lambda_7 (\rho^+ \eta)(\eta^+ \rho) \nonumber
+ \lambda_8 (\chi^+ \eta)(\eta^+ \chi) + \lambda_9 (\rho^+
\chi)(\chi^+ \rho) +
 \lambda_{10} (\chi^+ \eta + \eta^+ \chi)^2.
\label{pot}
\end{eqnarray}

One can then rewrite the expansion of the scalar fields
that  acquire a finite VEV:

\begin{equation}
  \eta^o = \frac{1}{ \sqrt{2}}\left(v +
\xi_\eta + i \zeta_\eta\right) ; \ \rho^o =  \frac{1}{
\sqrt{2}}\left( u + \xi_\rho + i \zeta_\rho\right);\ \chi^o =
\frac{1}{ \sqrt{2}}\left( w + \xi_\chi + i \zeta_\chi\right).
\label{exp1}
\end{equation}

 For the  prime neutral fields which do
not have a VEV, there are  analogous expressions: 
\begin{equation}
 \eta'^{o} = \frac{1}{
\sqrt{2}}\left( \xi'_\eta + i \zeta'_\eta\right) ; \ \chi'^{o} =
\frac{1}{ \sqrt{2}}\left( \xi'_\chi + i \zeta'_\chi\right).
\label{exp2} 
\end{equation}

In Ref.~[\refcite{ll}] the neutral scalars were studied and it was determined that they  can be both stable and  can satisfy the criteria for self-interacting dark matter.\cite{ss}  In addition, it was found that  that such dark matter particles do not overpopulate the universe within the other parameter  bounds on the model.
It was also shown\cite{ll} through a direct calculation that the Higgs scalar $h_0$ and $H_3^0$ can, in principle, satisfy the necessary criteria. Remarkably these do not interact directly with any SM field except for the standard Higgs $H_1^0$. However, the $h_0$  was favored \cite{ll} since it is easier to obtain a large scattering cross section for it, relative to $H_3^0$ for a convenient choice of the parameters.

In other singlet models\cite{Burgess2001a,Bento2000a, Holz01, Silvera85,McDonald02}
an extra symmetry was imposed to account for the stability of the dark matter.  However in the model of Ref.~[\refcite{ll}] the decay of the $h_0$ scalar is automatically forbidden to all orders of the perturbative expansion. This is because of two features: (i) this scalar comes from the triplet $\chi$, the one that induces spontaneous symmetry breaking of the 3-3-1 model to the standard model. Therefore, the SM fermions and the standard gauge bosons cannot couple with the $h^0$ ; and (ii)  the $h^0$ scalar comes from the imaginary part of the Higgs triplet $\chi$. The imaginary parts of $\eta$ and $\rho$ are pure massless Goldstone bosons. Therefore, there are no physical scalar fields that can mix with the $h^0$. Hence, the only interactions of $h^0$ come from the scalar potential.  They are the $H_3^0 h^0 h^0$ and $H_1^0 h^0 h^0$ couplings. The latter coupling has a strength of ${2i (-a_5v^2 + a_6u^2)}/{v_W}  2i \Theta$. 

It was concluded in Ref.~[\refcite{ll}] that  if $v\sim u\sim(100 \sim 200)$ GeV and $-1\leqslant a_5 \sim a_6 \leqslant 1$, the $h^0$ can interact only weakly with ordinary matter through the Higgs boson of the standard model $H_1^0$. 
The  quartic interaction for scattering is $h^0h^0h^0h^0$, with a strength of  $-ia_3$. Other quartic interactions evolving $h^0$ and other neutral scalars are proportional to $w^{-1}$ and can be neglected. The cross section of the process $h^0h^0 - h^0h^0 $ via the quartic interaction is 
\begin{equation}
\sigma = \frac{a_3^2}{64\pi m^2_h}~~.
\end{equation}
The contribution of the tri-linear interactions via $H_1^0$ and 
$H_3^0$ exchange are negligible. It has also been shown in Ref.~[\refcite{ll}]   that there are no other contributions to the process involving the exchange of vector or scalar bosons.

 \section{Constraints on 3-3-1 dark matter} 
General constraints on SIDM were deduced in Ref.~[\refcite{ss}].   Any self-interacting dark matter candidate should have mean free path $\Lambda={1}/{n\sigma}$ in the range 1 kpc $\ll 1$ Mpc, 
where 
$n = {\rho}/{m_h}$ is the number density of the $h_0$ scalar and $\rho$ is its density at the solar radius.\cite{moore,Klypin89} 
 
 Therefore, one can  obtain the required Spergel- Steinhardt bound,\cite{ss} of $2 \times 10^3$ GeV$^{3} \leqslant {\sigma}/{m_h} \leqslant 3 \times 10^4$ GeV$^{-3}$ with parameters $a_3 = -1, - 0.208 \times 10^{-7} GeV \leqslant f \leqslant
-0.112 \times10^{-6}$ GeV, $w=1000$ GeV, $u=195$ GeV and $\rho = 0.4$ GeV cm$^{-3}$.

With this set of parameter values, the constraint on the SIDM mass is  5.5 MeV $\leqslant m_h \leqslant$ 29 MeV. 
This means that the 3-3-1 SIDM dark-matter particle is non-relativistic during  the decoupling era
($T \sim 1$ eV) and, for a SM Higgs boson
mass\cite{pdg} of $ 125$ GeV, it is produced by
 thermal equilibrium $H^0 H^0 \leftrightarrows  h^0h^0$ with  the standard Higgs scalar.

\subsection{Relic density of the 3-3-1 SIDM}
The number density of the $h$  from the decaying of thermal equilibrium $H^0$ 
scalar can be obtained from a  solution to the Boltzmann equation.  For our purposes this can be reduced to \cite{Koln90}
\begin{equation}
\frac{dn_h}{dt} + 3Hn_h = <\Gamma_{H^0}> n_{H^0{eq}},
\end{equation}
where $n_{H^0{eq}}$ is the thermal equilibrium density of the Higgs scalar $H^0$;  $H$ is the expansion rate, 
 and $\Gamma_H$ is the decay rate for the $H^0$ scalar  with  energy $E$. 
 
 The thermal equilibrium density of the standard $H^{0}$ at temperature $T$  is given by \cite{Koln90} 
\begin{equation}
n_{H^0{eq}}=\frac{1}{2\pi^2} \int_{m_H}^{\infty} \frac{E \sqrt{E^2-m_H^2}dE}{e^{E/T}-1}   ~~.
\end{equation}
 Adopting  the condition that the temperature is lower than the electroweak phase transition temperature $T_{EW} \geqslant 1.5 m_H $. The thermal average of the decay rate for $H_0$ is given by:
\begin{equation}
\Gamma_{H^0} =\frac{\lambda^2_hv^2}{16\pi E} =  \frac{\alpha   \lambda_h^2 v^2 T^2}{32 \pi^2 n_H  } e^{m_H/T}
\end{equation}
where $\lambda_h$ is the scalar self coupling, and T is the background cosmic temperature, $\alpha \approx 1.87$ is an integration parameter. Defining  $\beta \equiv  {n_h}/{T^3}$  in the radiation dominated era  the evolution of the Boltzmann equation  can be written as

\begin{equation}
\frac{d \beta}{dT} = - \frac{\Gamma \beta}{KT^3} = -\frac{\alpha \lambda_h^2 v^2}{32\pi^3 T^4 K e^{m_1/T}}  ~~,
\end{equation}

where $K^2 ={4\pi^3 g(T)}/{45m^2_{Pl}}$, $m_{Pl} = 1.2 \times 10^{19}$ GeV is the Planck mass,  and $g(T)=g_B +7 {g_F}/{8}=136.25$  \cite{Koln90}  where $g_B$  and $g_F$ are the relativistic bosonic and fermionic degrees of freedom, respectively.  One can take $T = m_h$ since this regime gives the larger contribution to $\beta$. Then,
\begin{equation}
\beta = \frac{\alpha \lambda_h^2v^2}{16\pi^3 Km_1^3}~~.
\end{equation}
so that the cosmic density of the $h_0$ scalar becomes
\begin{equation}
\Omega_h = 2g(T_\gamma)T^3_\gamma \frac{m_h\beta}{\rho_c g(T)} ~~,
\label{1}
\end{equation}
where $T_\gamma=2.4 \times10^{-4}eV$ is the present photon temperature, $g(T_\gamma)= 2$ is the photon degree of freedom and $\rho_c = 7.5 \times 10^{-47} h^2$, with h=0.71, being the critical density of the universe, $g(T)= g_B+\frac{7}{8}g_F = 136.25$ \cite{ft} , $g_B$ and $g_F$ denote the
number of relativistic bosonic and fermionic degrees of freedom respectively.

As an illustration, it was shown in Ref.~[\refcite{ll}] the parameters  $m_h$ = 7.75 MeV, v = 174 GeV, $a_5$ = 0.65, -$a_6$ = 0.38 ,  and $m_1$ = 150 GeV lead to  $\Omega_h = 0.3$. Hence, without imposing any new fields or symmetries, the 3-3-1 model possesses a scalar field that can satisfy all the desired properties for the self-interacting dark matter while producing  the correct relic density.  Moreover, the relic density implies a mass  range of  $5.5 MeV \leqslant m_h \leqslant 29 MeV$
within the bounds of the model.\cite{ll}

\subsection{Constraint from Galactic Structure}

In this  work we expand on the work or Ref.~[\refcite{ll}] and consider the simulations of the detailed relation between the SIDM density and  galactic structure as well as the Cosmic Microwave Spectrum power spectrum as a better means to constraint the 3-3-1 SIDM model. 
Models for self interacting dark matter should have a core size less than 2 kpc on the scale of dwarf galaxies.\cite{Burkert95}
For the scalar $h_0$  candidate for self interacting dark matter, we allow a range for the density of the SIDM of 0.11 to 0.5  

 For our simulations, we utilize both  the AMR ENZO  code\cite{enzo}  and the N-body simulation code of Koda \& Shapiro\cite{Koda11}
 
 The global properties of a number of runs based upon  typical dwarf spheroidals are given in Table \ref{ta1}. 
 
 The Friedmann equation in the model including  SIDM is 
\begin{equation}
 H^2=\frac{8 \pi G}{3} (\rho_{\Lambda} + \rho_{SM}+\rho_{DM})
\end{equation}
 where $\rho_\lambda$ is the dark energy density, $\rho_{SM}$ is the density dark matter non-interacting, $\rho_{SIDM}$ is the density of dark matter interacting, 
Within this expansion we utilize the Monte Carlo algorithm of Ref.~\refcite{Koda11} to model  the non-gravitational scattering of DM particles by other DM particles, within a gravitational N-body method. 
\begin{figure*}
\begin{center}
\includegraphics[height=8cm]{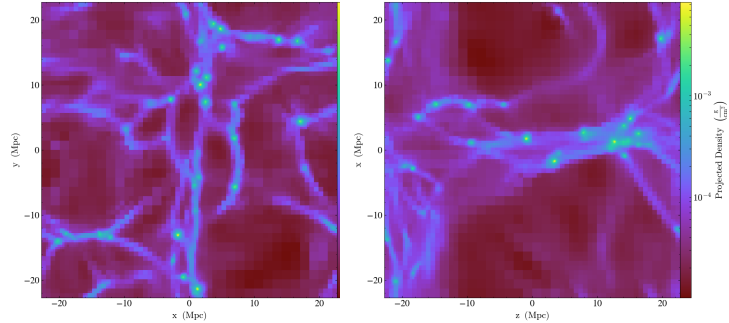}
\caption{Simulation of density contours of the self-interacting dark matter (left) and cold dark matter (right)}
\label{fig:2}
\end{center}
\end{figure*}

 
 We simulated the formation of galactic dark matter halo with $10^{10} M_\odot $  with varying cross section for self interacting dark matter scattering. We study the galaxy formation with stellar masses in self interacting dark matter (SIDM) with with cross section over mass in the range of  $3.7$ to $5.2 $ cm$^2$ g$^{-1}$.
 The Lambda Cold Dark Matter model considers the DM to be  non relativistic and collisionless. However self interacting dark matter has a weak interaction with the Higgs Boson in the Standard Model of particle physics. 
 
We start with isolated halo galaxies with a stellar mass of $M_{star} = 1.4 \times 10^{11}$ M$_{\odot} $ and temperature  $T=10^4$K with in a box size of $50$ Mpc $h^{-1}$.  
For the Cold Dark Matter run we use the initial radial density profile\cite{Navarro97} as defined in Eq.~\ref{eq:1} and illustrated in Figure \ref{fig:1}.
The initial profile for the SIDM simulations was given\cite{Burkert95}  by Eq.~(\ref{eq:2})  and illustrated in Figure \ref{fig:1}.

\begin{table}[h]
\tbl{Simulation parameters.}
 {\begin{tabular}{@{}ccccc@{}}\toprule
 Name & Volume ($h^{-1}$ Mpc) & $N_p$& $m_p (h^{-1} M_{\bigodot})$ &   $\frac{\sigma}{m} (cm^2 g^{-1})$\\ 
\colrule
 CDM & 50 & $512^3$ & Particles mass &  0 \\ 
SIDM-3.5 & 50 & $512^3$  & Particles mass &  3.5 \\  
SIDM-5.5 & 50 & $512^3$  & Particles mass &  5.5 \\ 
\botrule
\end{tabular}\label{ta1} }
\end{table}

\subsection{Constraints on SIDM Parameters}

Based upon our fits to observed dwarf spheroidal galaxies, we have deduced  the acceptable range of SIDM parameters.  The range of mass for SIDM in the MeV range with a mean free path  and total cross section over mass from 0.1 $cm^2 g^{-1}$ to $100 cm^2 g^{-1}.$ These simulations are close to the observations in the inner DM halo and can solve the core-cusp and the missing satellite problems of the $\Lambda$CDM model. In our model the cross section over mass$\frac{\sigma}{m}$ is in the range of 4 to 5 $cm^2 g^{-1}$. In the large scale structure there is no difference with normal CDM but in small scale structure the core  is consistent with all of the observational constraints (see Figure 2.) In general, SIDM would make no difference from $\Lambda$CDM on large scales, however, individual galaxies would have dense spherical cores and a higher velocity dispersion.

\begin{figure}[htp]
\begin{center}
\includegraphics[height=6cm]{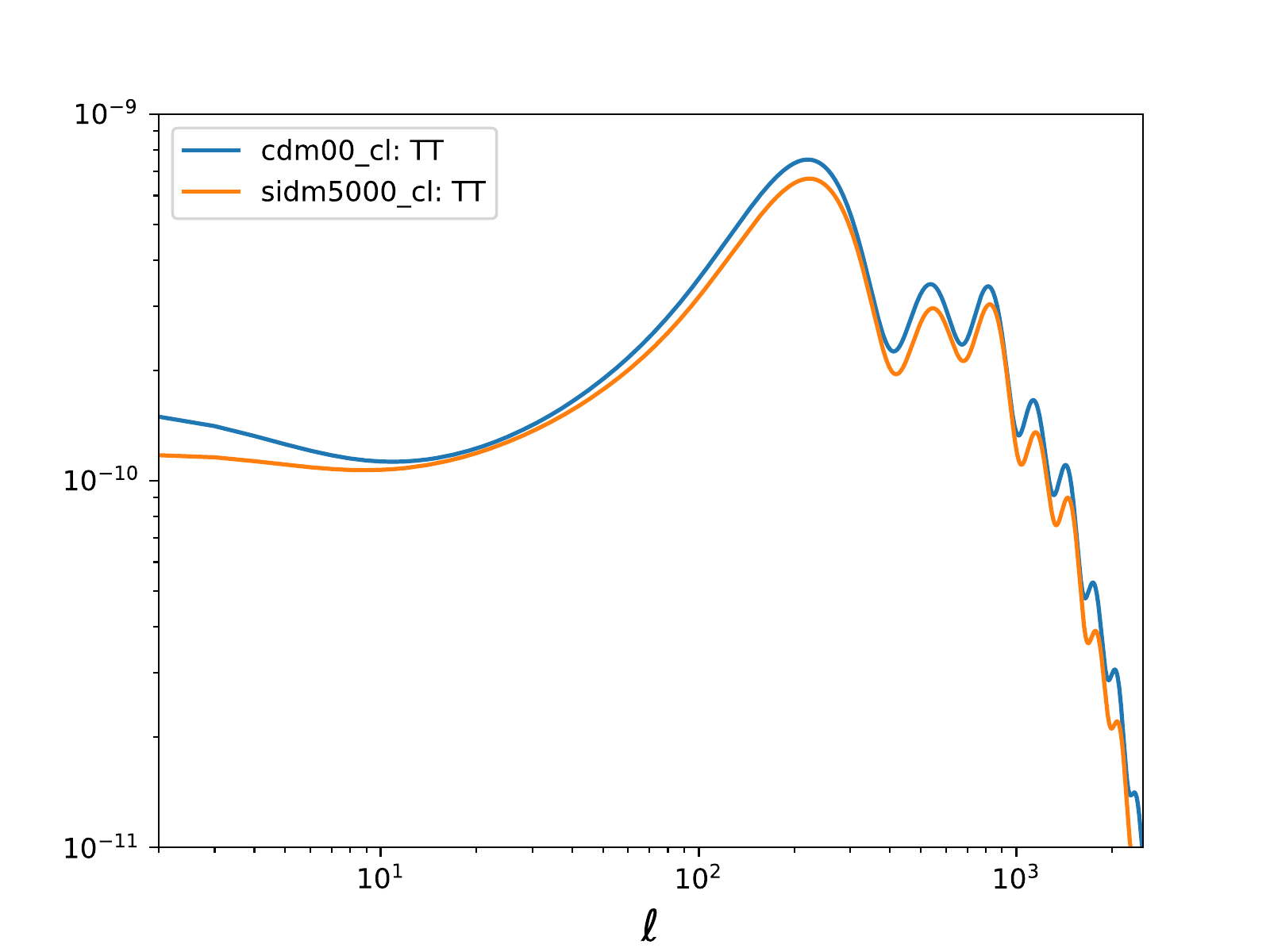}
\caption{The power spectrum of the self-interacting dark matter is lower than cold dark matter in TT mode with multipole l less than $10^3$ }
\label{fig:4}
\end{center}
\end{figure}

\begin{figure}[htp]
\begin{center}
\includegraphics[height=6cm]{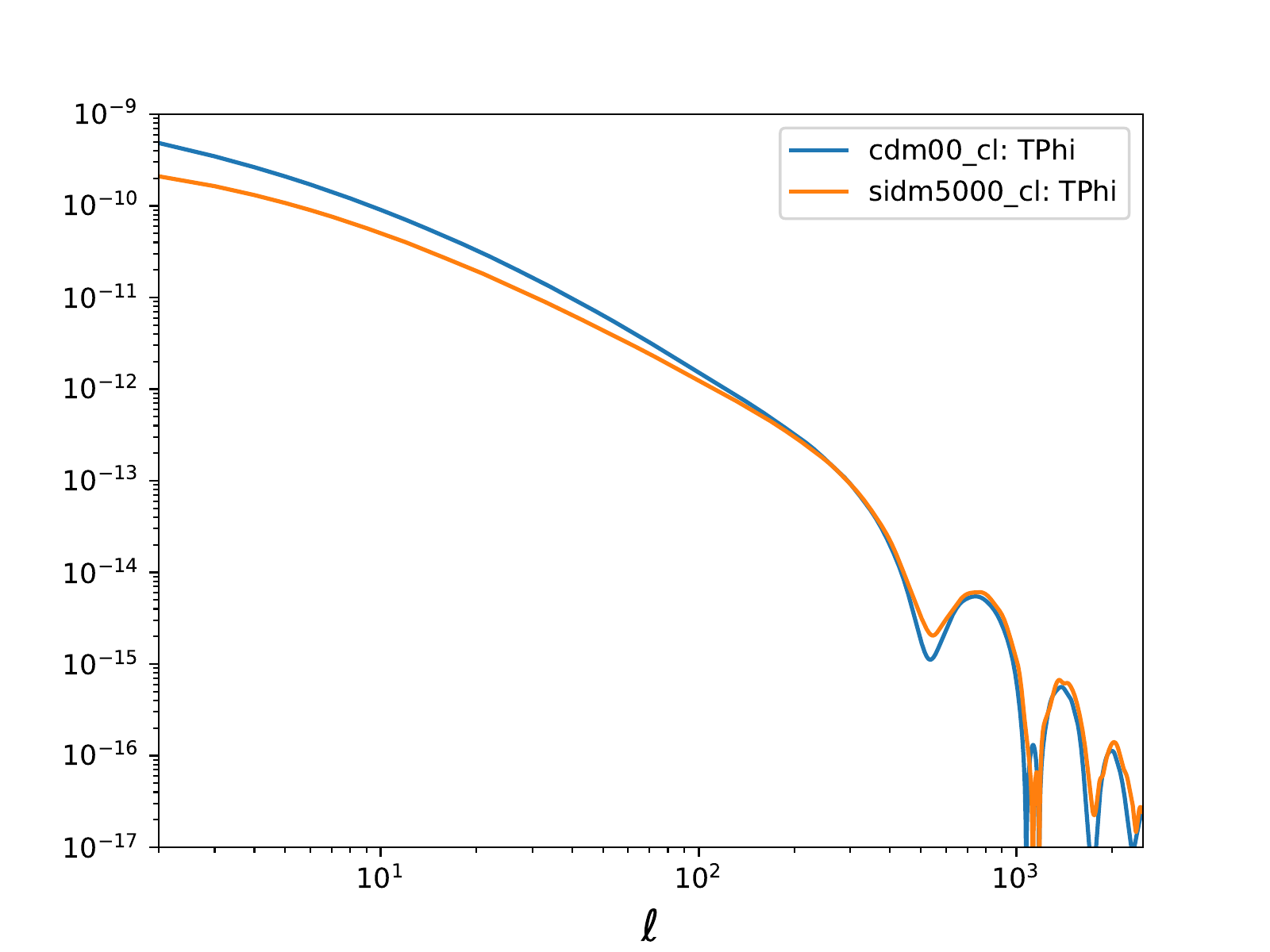}
\caption{The power spectrum of the self interacting dark matter is lower than cold dark matter in Tphi mode with multipole l less than $10^1$ }
\label{fig:5}
\end{center}
\end{figure}

\begin{figure}[htp]
\begin{center}
\includegraphics[height=6cm]{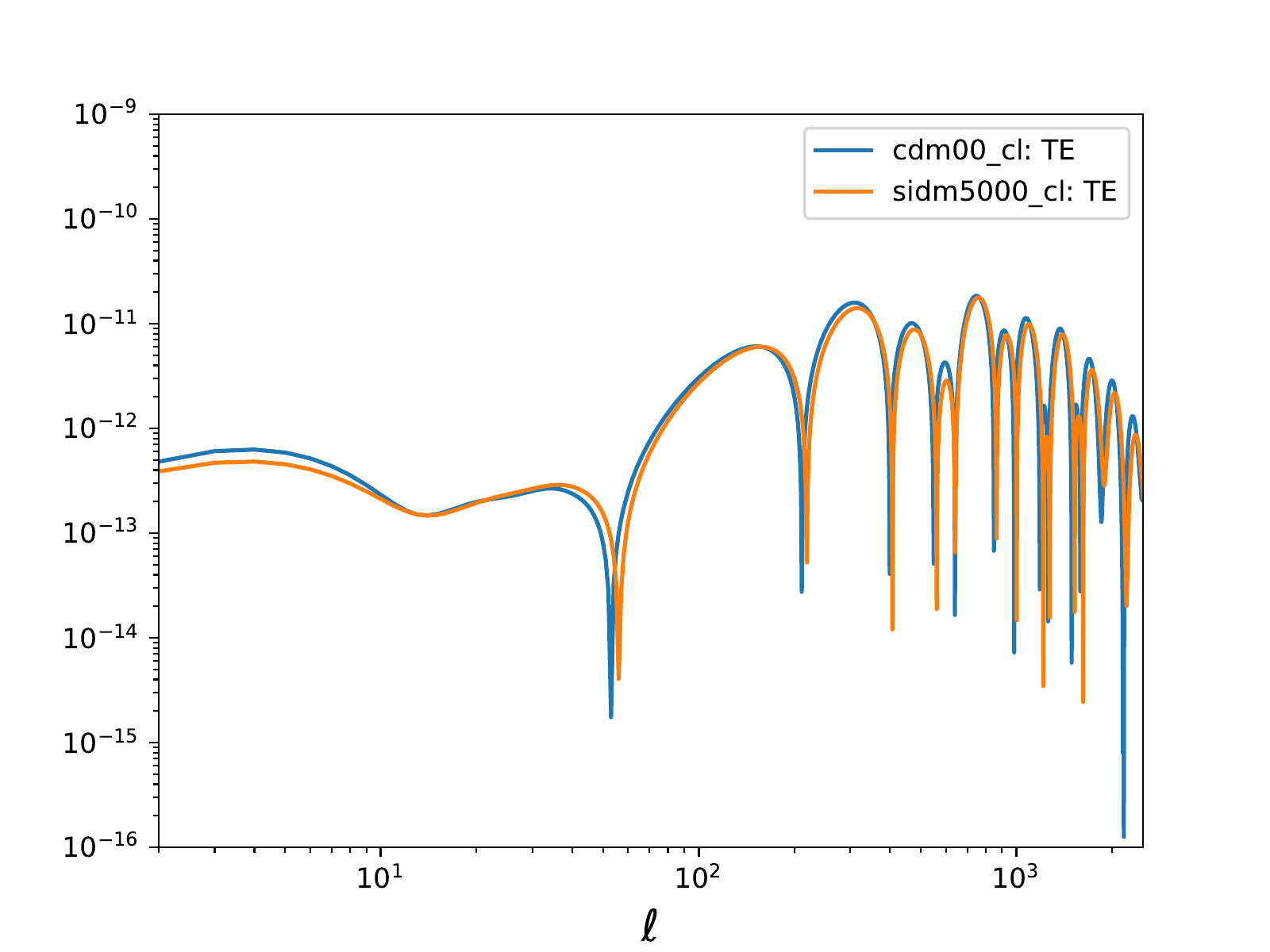}
\caption{The power spectrum of the self interacting dark matter in mode with multipole l less than $10^2$}
\label{fig:6}
\end{center}
\end{figure}
 With the hydrodynamic simulation, we can see the effectively the small scale structure of the dark matter. 
We have used the Planck temperature and WMAP polarization data to constrain the self interacting dark matter (see Figures 3-5.) We use cosmological model with parameters\cite{Plank collaboration 2018} $\Omega_\Lambda =0.6894, \Omega_m=0.3147, \Omega_b=0.022447, n_s=0.96824, A_s=2.173, h=0.673, \sigma_8 = 0.805$ and code {\it CLASS}. \cite{class} We see that the power spectrum of self interesting dark matter is smaller than the power spectrum of dark matter in the region of small multipoles, and much smaller for the multipoles less than 10.
\section{Conclusion }
The 3-1-1 model remains as a viable candidate for SIDM.  Here we have reviewed the attractive features of the model and propose new constraints on parameters based upon fits to the observed dark matter profiles of dwarf spheroidal galaxies.  We find that the cross section over mass is in the range from  3.7 - 5.2  $cm^2 g^{-1}$  for the data of dwarf spheroidal galaxies of the Milky Way \cite{dwraftgalaxy}
We note, however, that there are also constraints on SIDM based upon its effect on the CMB.  Also, it has been noted\cite{Men} that self-interacting models lead to a spherical halo in centers of clusters, which may not be in agreement with the ellipsoidal
centers indicated by strong gravitational lensing\cite{yoh00} and by Chandra
observations\cite{buote}. Nevertheless, self-interacting dark matter models remain  well motivated as
a solution to the core-cusp problem.
\section{Acknowledgments}
This work was supported by the
U.S. Department of Energy under Nuclear Theory grant DEFG02-95-ER40934. Lan Q. Nguyen gratefully acknowledges support from the Simons Center for Geometry and Physics, Stony Brook University at which some of the research for this paper was performed.

\end{document}